\pdfoutput=1
\documentclass[12pt]{article}

\usepackage[margin=1in]{geometry}  % set the margins to 1in on all sides
\usepackage{graphicx}              % to include figures
\usepackage{amsmath}               % great math stuff
\usepackage{amsfonts}              % for blackboard bold, etc
\usepackage{amsthm}                % better theorem environments
\usepackage{tabularx}
\usepackage{program}

\usepackage[hang,small,bf]{caption}
\newtheorem{defin}{Definition}[section]
\newtheorem{rem}{Remark}[section]
\newtheorem{thm}{Theorem}[section]
\newtheorem{lem}{Lemma}[section]

\newtheorem{cor}{Corollary}[section]

\newtheorem{assum}{Assumption}[section]

\newcommand{\keywords}[1]{\par\addvspace\baselineskip\noindent\enspace\ignorespaces#1}

\begin{document}

\nocite{*}

\title{ Introduction into ``Local Correlation" Modelling}

\author{ Alex Langnau \\ \small{Allianz SE and LMU Munich}}

\maketitle
\vspace{2cm}

\begin{abstract}
In this paper we provide evidence that financial option markets for equity indices give rise to non-trivial dependency structures between its constituents. Thus, if the individual constituent distributions of an equity index are inferred from the single-stock option markets and combined via a Gaussian copula, for example, one fails to explain the steepness of the observed volatility skew of the index. Intuitively, index option prices are encoding higher correlations in cases where the option is particularly sensitive to stress scenarios of the market. As a result, more complex dependency structures emerge than the ones described by Gaussian copulas or (state-independent) linear correlation structures.
In this paper we ``decode" the index option market and extract this correlation information in order to extend the multi-asset version of Dupire's ``local volatility" model by making correlations a dynamic variable of the market. A ``local correlation" model (LCM) is introduced for the pricing of multi-asset derivatives. %We show how consistency with the index volatility data can be achieved by construction.
LCM achieves consistency with both the constituent- and index option markets by construction while preserving the efficiency and easy implementation of Dupire's model.


\end{abstract}

\pagebreak

\section{Introduction}
In 2008, the crisis in credit markets led to a spike in correlations that resulted in large losses for several single-stock exotic trading books. As most market participants did not anticipate the crises, one is tempted to also put these equity losses under the category of unpredictable and unforeseen ``bad" events. We do not fully subscribe to this view.
\newline \linebreak
In this paper we argue that, at least to some extent, the losses could have been alleviated if models had incorporated non-trivial correlation information, already encoded in the index option market, properly into the pricing of multi-asset derivatives. This is because the index volatility market already implies rising correlations in downturn scenarios.

Most investment banks utilise a Gaussian copula for the pricing of multi-asset exotic European payoffs. For more involved payoffs a multi-asset version of Dupire's local volatility model has become the market standard for the pricing and hedging of those products. The model infers a local spot-dependent volatility surface from option prices for each asset and combines them via a (constant) correlation matrix between its Wiener increments. However even when correlations are assumed time-dependent, the model does not incorporate all dependency information inherent in the derivatives market properly. In particular with downward scenarios, index options imply correlations that can be substantially higher compared to the ``at-the-money" case. As a result the cost of short cross-gamma position hedging could be priced in advance, more appropriately, by means of a better model.

In chapter 2 of this paper we take a snapshot of the options market and deduce the distribution of the Dow Jones Euro Stoxx 50 (.STOXX50E) and DAX (.GDAXI) indices together with all the composite distributions. In both cases we show that only half of the volatility skew of the index is explained by the skew of the index constituents. We conclude that (term) correlation effectively depends on the strike of the option which substantially impacts the distribution of the index on the downside.

It is important that multi-asset derivative models incorporate these effects even if a product is only sensitive to a subset of stocks in the index. This is because the picture remains fundamentally the same: stocks correlate more strongly in a bear market, when investment behaviour is driven by fear, than during bullish times. As a result portfolio managers may find themselves less diversified, even when the composition of their portfolio stays the same. As far as hedging of derivatives is concerned, an increase in correlations has the unpleasant effect of increasing the cost of managing short cross-gamma positions. This arises as a result of the non-linear behaviour of (multi-asset) options' prices and manifests itself in the fact that the option's sensitivity to one asset increases with the value of a second decreasing. During the credit crunch for example, the Nikkei often finished down for the day creating a long delta position in EuroStoxx 50 for a hedging book in London, even though the market there was closed. As a result, the trader would need to immediately sell EuroStoxx 50 at the open. However EuroStoxx 50 would  be likely to open down as well because of the large positive correlation between the two assets. This gives rise to systematic losses during periods when the correlation exceeds expectations. 
\linebreak \linebreak
The goal of this paper is to develop a model that incorporates higher correlations in downward scenarios by ``decoding" the state-dependent correlation information of the index option market. One expects that such a model anticipates losses such as those described above at the inception of the trade and adjusts the pricing accordingly. In addition one would expect more accurate deltas in this case.

In order to make LCM a model that can be useful for practical applications, one needs to provide a robust as well as efficient implementation for the underlying algorithms. This is the strenght of Dupire's local volatility model which is simple to implement, numerically sufficiently efficient and provides consistency with the volatility skew of individual option markets at the same time. The goal of this paper is to extend this methodology to the multi-asset case by providing consistency with the  index option market in addition to the individual option markets. Note that, similar to the Dupire framework, LCM provides a particular way of achieving consistency with the option markets, however this solution is by no means unique. Chapter 7 provides
alternative solutions to the same problem which allows one to quantify "residual exotic correlation risk"'.
\newline

Work in this direction has also been pursued by others. See for example \cite{Langnau2006},\cite{cit1},\cite{cit2},\cite{cit3},\cite{cit4}.
\newline

Chapter 3 lays down the ground work for the ``local correlation" model (LCM) that is introduced in Chapter 4. A parameterisation of correlations is introduced that preserves the positive semi-definiteness and displays some notion of strictly increasing correlation matrices. Chapter 5 provides simulation results of LCM and provides explicit results for the strike dependence of correlation. Chapter 6 deals with some implementation details and chapter 7 discusses residual correlation risk. Chapter 8 concudes our analysis.

\pagebreak

\section{``Decoding" the Index Options Market}
\label{sect:decoding}
In this chapter we take a snapshot of the market as of 31 July 2009 and try to explain the observed index option prices of EuroStoxx 50 and DAX in terms of their constituent distributions. One finds that the skewness of the individual distributions fails to fully explain the volatility skew of the index basket itself. Hence, the index volatility market encodes non-trivial dependencies between its constituents. Let $I_T$  denote the value of the index at time $T$. The price of a call option on the index with maturity $T$ and strike $K$ is given by $C(T,K)$. Note that

\begin{align}
\label{form:rnpricing}
C(T,K) = {\text {\textit{ DF}}}\,E^Q((I_T-K)^+) = {\textit{ DF}} \int_0^{\infty} {dI} \hspace{5mm}prob(T,I) (I-K)^+
\end{align}
where DF is the discount factor to maturity and $E^Q$ denotes the expectation value under the forward adjusted measure $Q$. Thus, Eq.\ref{form:rnpricing} relates option prices to the (forward-adjusted) implied density $prob(T,I_T)$ that is inferred by the option market with maturity T.
Assume that for a given maturity T all strikes are traded. In this case Eq.\ref{form:rnpricing} can easily be inverted with the result
\begin{align}
\label{form:distr}
\frac{1}{{\text {\small{ DF}}}}\frac{\partial^2 C(T,K)}{\partial K^2} = prob(T,K).
\end{align}
Let us denote by $S_i(T)$, $i=1,..,n$, the price of constituent $i$ at time $T$. The index is generally defined as a weighted sum of its constituents e.g.
\begin{align}
I_t = \sum_{i=1}^n {\alpha_i \hspace{1mm} S_i(t)}.
\end{align}

%\pagebreak
Examples for basket weights are given in Table 1. For the DAX the weights are listed in Table 2.

\begin{table}
\begin{center}
\begin{scriptsize}

\label{decomp}

\begin{tabular}{|rr|rr|rr|rr|rr|}
\hline
\hline
Ticker & Wgt & Ticker & Wgt &Ticker & Wgt &Ticker & Wgt &Ticker & Wgt\\
\hline
AEGN.AS	&2.67&AIRP.PA	&0.50&ALVG.DE	&0.86&ALSO.PA	&0.38&ISPA.AS	&1.76\\
GASI.MI	&2.26&AXAF.PA	&3.17&BBVA.MC	&7.12&SAN.MC	&15.50&BASF.DE	&1.75\\
BAYG.DE	&1.57&BNPP.PA	&1.66&CARR.PA	&1.16&SGOB.PA	&0.76&CAGR.PA	&1.93\\
DAIGn.DE &1.69&DBKGn.DE &1.18&DB1Gn.DE &0.37&DTEGn.DE &5.66&EONGn.DE &3.80\\
ENEI.MI	&12.11&ENI.MI	&4.63&FOR.BR	&4.16&FTE.PA	&3.61&GSZ.PA	&2.53\\
DANO.PA	&1.14&IBE.MC	&6.54&ING.AS	&3.95&ISP.MI	&18.33&PHG.AS	&1.85\\
OREP.PA	&0.45&LVMH.PA	&0.49&MUVGn.DE &0.38&NOK1V.HE &7.12&RENA.PA	&0.38\\
REP.MC	&1.50&RWEG.DE	&0.77&SASY.PA	&1.99&SAPG.DE	&1.66&SCHN.PA	&0.49\\
SIEGn.DE&1.55&SOGN.PA	&1.10&TLIT.MI	&19.43&TEF.MC	&7.79&TOTF.PA	&3.99\\
CRDI.MI	&30.25&UNc.AS	&2.95&SGEF.PA	&0.94&VIV.PA	&2.33&VOWG.DE	&0.16\\
\hline
\hline
\end{tabular}

 \vspace{5mm}

\caption{Decomposition of EuroStoxx 50 on the 31 July 2009. The first column displays the (Reuters) names of the assets followed by the weight in the index.}

\end{scriptsize}
\end{center}
\end{table}
\normalsize

\begin{table}
\begin{center}
\begin{scriptsize} 
\label{decompdax}

\begin{tabular}{|rr|rr|rr|}
\hline
\hline
Ticker & Wgt & Ticker & Wgt &Ticker & Wgt\\
\hline
ADSG.DE	&2.20&ALVG.DE	&5.16&BASF.DE	&10.46\\
BMWG.DE	&3.66&BAYG.DE	&9.41&BEIG.DE	&0.93\\
CBKG.DE	&8.20&DAIGn.DE	&9.72&DBKGn.DE	&7.04\\
DB1Gn.DE	&2.12&DPWGn.DE	&9.57&DTEGn.DE	&33.91\\
EONGn.DE	&18.76&FMEG.DE	&2.14&FREGp.DE	&0.92\\
HNRGn.DE	&0.68&HNKGp.DE	&2.03&SDFG.DE	&1.40\\
LING.DE	&1.92&LHAG.DE	&5.21&MANG.DE	&1.13\\
MRCG.DE	&0.74&MEOG.DE	&1.26&MUVGn.DE &2.25\\
RWEG.DE	&4.71&SZGG.DE	&0.43&SAPG.DE	&9.63\\
SIEGn.DE	&9.28&TKAG.DE	&3.79&VOWG.DE	&0.98\\
\hline
\hline
					
\end{tabular}

\caption{Decomposition of DAX on the 31 July 2009. The first column displays the (Reuters) names of the assets followed by the weight in the index.}

\end{scriptsize}
\end{center}
\end{table}

Theoretically the decomposition could change during the life of the option. However these events are relatively rare so that we ignore potential effects for the pricing in this paper.
Let us denote by $C_i(K,T)$ the options prices of the $i$-th constituent. Similar to before, these prices are observed in the market so that individual distribution can be extracted according to
\begin{align}
\label{form:individdistr}
\frac{1}{DF}\frac{\partial^2 C_i(T,K)}{\partial K^2} = prob_{i}(T,K).
\end{align}
The integration of Eq.\ref{form:distr} and Eq.\ref{form:individdistr} yields for the cumulative distributions
\begin{align}
\label{form:cumdistr}
P(I_T < K) &= 1+\frac{1}{DF}\frac{\partial C(T,K)}{\partial K}\\
P_i(S_i(T) < K) &= 1+\frac{1}{DF}\frac{\partial C_i(T,K)}{\partial K}.
\end{align}
Let $\omega$ denote a $n$-dimensional standard normal correlated Gaussian variate with
\begin{align*}
\label{form:corrvariate}
E^Q(\omega_i \omega_j) &= \rho_{ij} \hspace{5mm} i \neq j
\end{align*}
\begin{align}
E^Q(\omega_i) &= 0
\end{align}
\begin{align*}
E^Q(\omega_i^2) &= 1
\end{align*}
for a given correlation matrix $\rho_{ij}$. If one assumes a Gaussian copula one can construct the joint distribution from the constituents by inversion of
\begin{align}
n_i \equiv N(\omega_i) = P_i(S_i(T) < K_i) \hspace{10mm} i=1,\ldots,n.
\end{align}
The resulting multivariate distribution is denoted by
$$ prob(T,S_1(T),S_2(T),\ldots,S_n(T)).$$
Hence, index option prices can be computed in two different ways:
\begin{enumerate}
\item	by means of Eq.\ref{form:rnpricing} representing the market quotes of the index option
\item	by means of multidimensional integration over its constituents, e.g.
\begin{align}
\widetilde{C}(K,T) = \text{DF}\int_0^{\infty} dS_1 ...dS_n \hspace{5mm} prob(T,S_1,S_2,\ldots,S_n)\left\{\sum_{k=1}^n \alpha_k S_k-K\right\}^+
\end{align}
\end{enumerate}
If a Gaussian copula was to represent an adequate picture for combining individual spot distributions one should find 
\begin{align}
C(K,T) = \widetilde{C}(K,T) \hspace{10mm} \forall{K,T}.
\end{align}

We have carried out this calculation, with some of the results depicted in Fig.1, Fig.2, Fig.3.
However instead of comparing option prices or distributions directly it is more meaningful to compare implied volatilities instead.

\begin{figure}
\begin{center}
\includegraphics[width=0.9\textwidth]{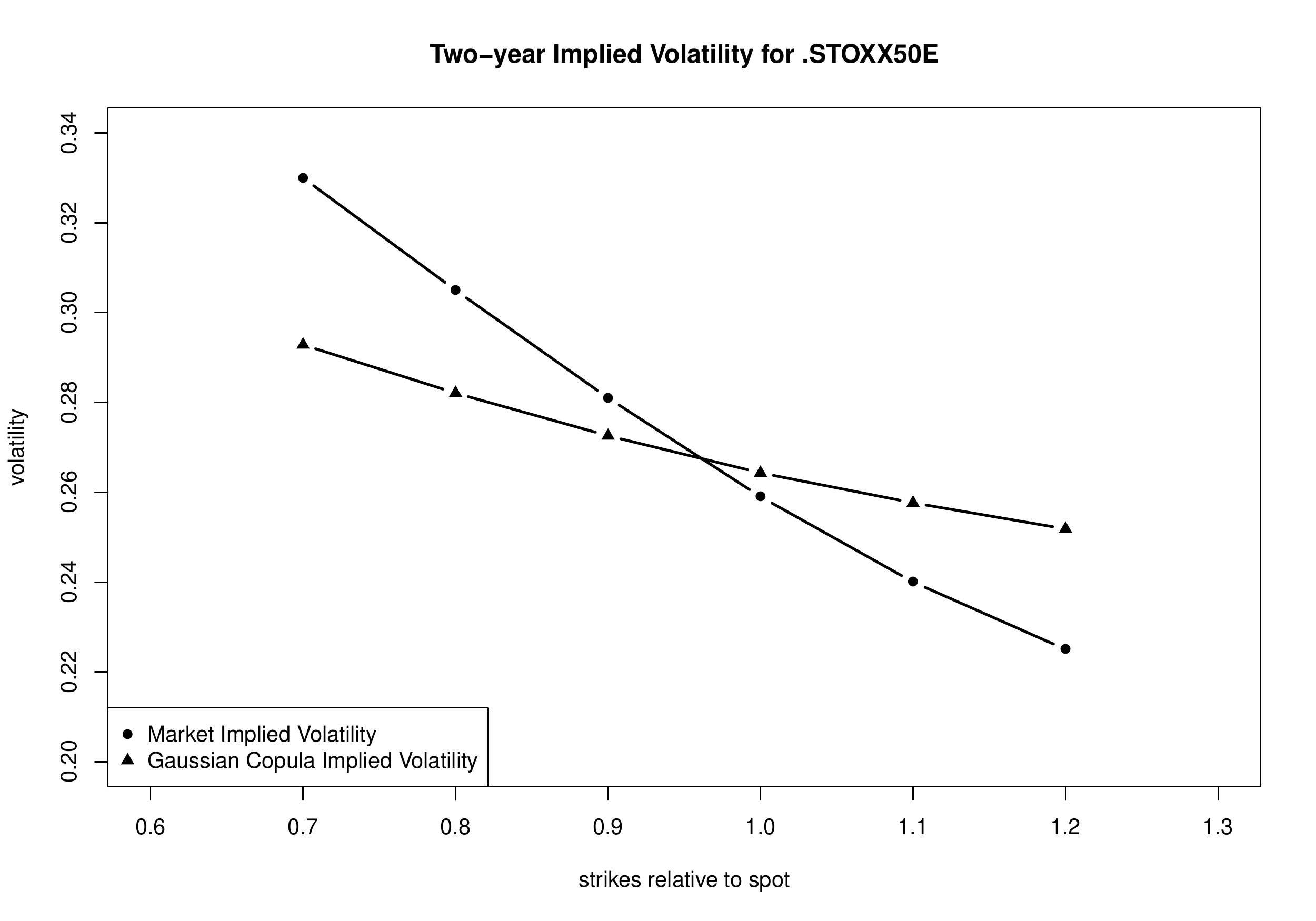}
\caption{Comparison of implied volatilities for the two year option market of EuroStoxx 50 between the market quotes and option prices obtained via multidimensional integration assuming a Gaussian copula. Only (roughly) half the skew is accounted for by the skew of the individual distributions.}

\end{center}
\end{figure}

Fig.1 shows that individual option distributions fail to explain the volatility skew of the index for both EuroStoxx 50. Only roughly half of the skew can be attributed to the skew of the individual assets themselves. Note that this inconsistency does not necessarily imply dispersion arbitrage opportunities between the index option market and the options markets of and the individual assets. It is more reasonable to conclude that the option market data suggests a type of dependency between individual stocks that is more complex than the one described by a Gaussian copula.

\begin{figure}
\begin{center}
\includegraphics[width=0.9\textwidth]{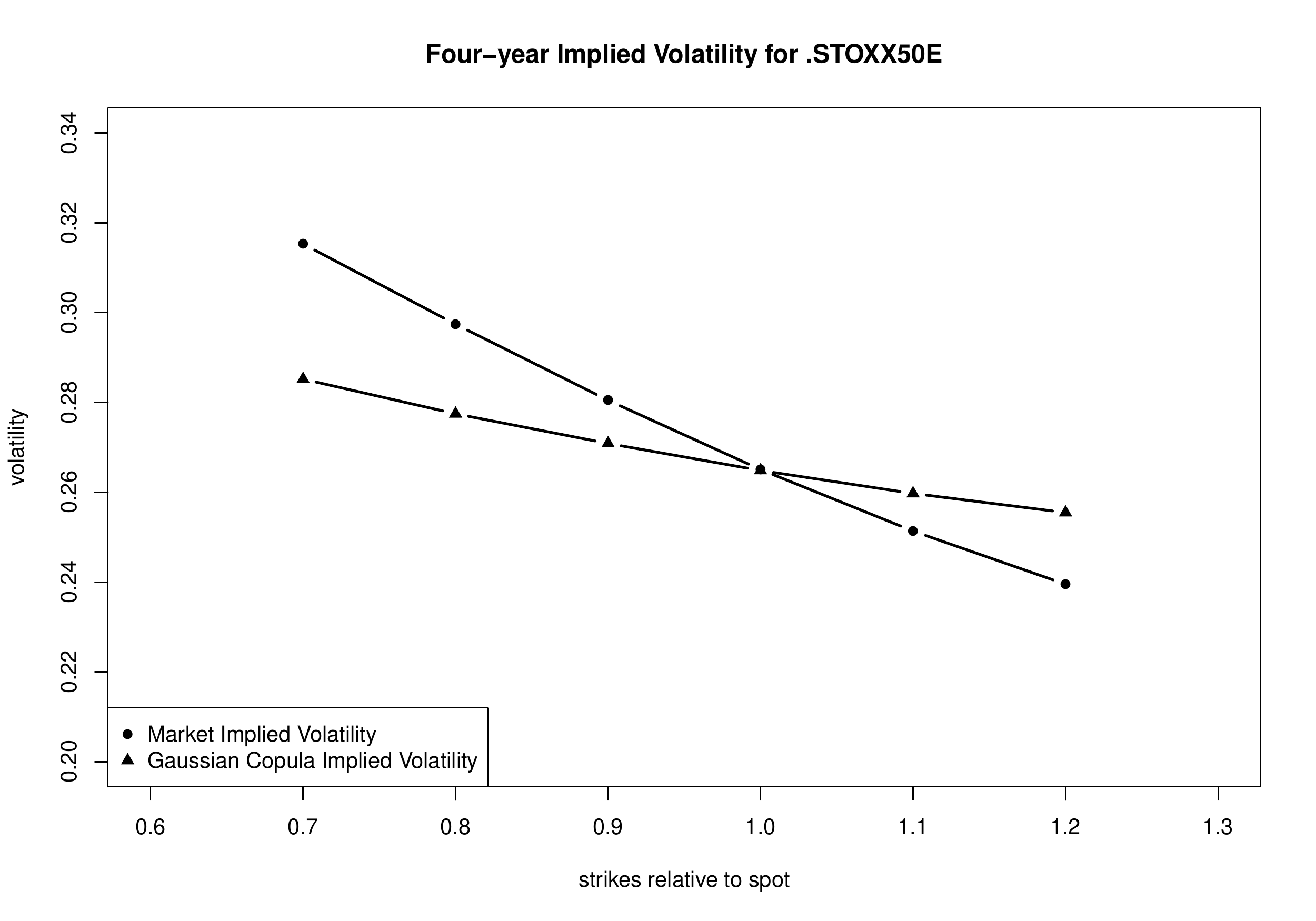}
\caption{Comparison of implied volatilities for the 4-year option market of EuroStoxx 50 and results obtained by integration of individual distributions that were combined by a Gaussian copula.}

\end{center}
\end{figure}

In order to study whether the effect is persistent in time Fig.2 displays the result of the 4-year market. The qualitative picture remains the same: Deviations from normality of individual log-returns cannot explain the non-normality of log-returns of the index.
\newline

\begin{figure}
\begin{center}
\includegraphics[width=0.9\textwidth]{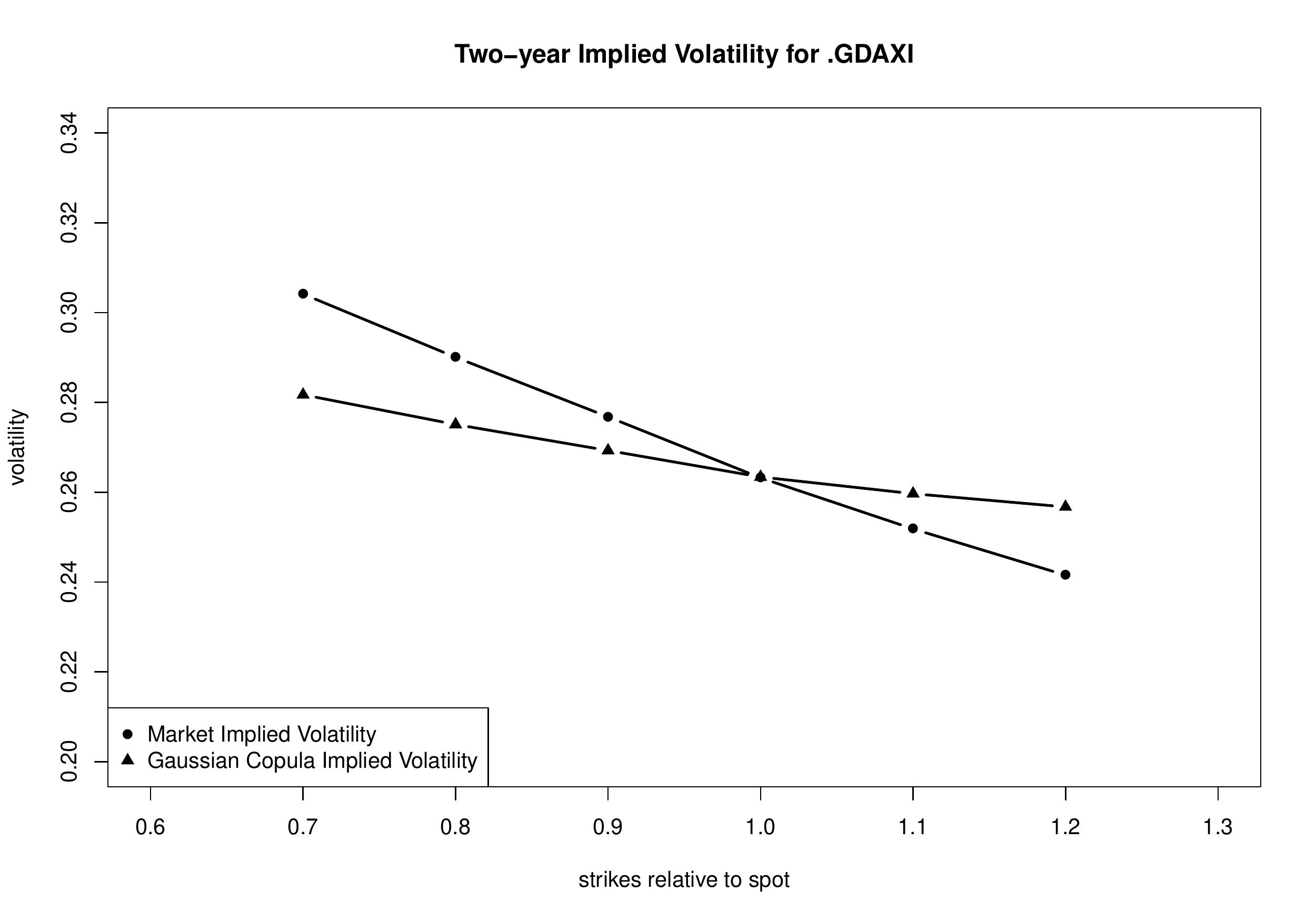}
\caption{ Comparison of implied volatilities for the 4-year option market of DAX and result obtained by integration of individual distributions that were combined by a Gaussian copula.}

\end{center}
\end{figure}

The results presented so far could represent an anomaly of EuroStoxx 50. However Fig.3 displays the result of calculations for the DAX that are qualitatively similar.

Fig.1,2,3 show that individual option distributions fail to explain the volatility skew of the index for both EuroStoxx 50 and DAX. Only roughly half of the skew can be attributed to the skew of the individual assets themselves. Also, other choices for the correlation matrix in Eq.\ref{form:corrvariate} fail to generate enough skew. Possible improvements by ``scanning" the space of correlation matrices would be marginal at best. Note that reasonable choices for correlation matrices should result in basket implied volatilities that match the index's implied volatility curve around the most liquid point, which typically occurs at a strike of about 100$\%$.

Thus, the option market data strongly suggests the behaviour of dependencies between individual stocks in the index are more complex than the one described by a Gaussian copula.

\pagebreak

\section{One parameter family of correlation matrices}
\label{chap:perturbations}

In this chapter we lay down the groundwork for the local correlation model (LCM) described in the next chapter. We define a family of perturbations of a given (center) correlation matrix that preserves all the relevant properties of the system, such as its positive semi-definiteness. The construction of this family is motivated by Poisson jump processes (common jumps generally increases the correlation between assets while uncorrelated jumps cause de-correlation) even though jumps do not play an explicit role in the LCM model. The important result of this chapter lies in the fact that a one-parameter family of correlations can be defined that allows one to continuously dial correlations in between the identity matrix and a correlation matrix that is identical to one, i.e.$\rho_{ij}=1$. 

\begin{defin}
\label{def:corrMatrix}
Let $V^{(n)}$ be a $n$-dimensional vector space in $R$ with a canonical basis given by $\left\{e_i: i=1,\ldots,n\right\}$. A linear operator that is represented by a $n\times n$  matrix $C$ in the basis $\left\{e_i\right\}$ is called a correlation matrix if
%\begin{enumerate}
\begin{itemize}
\item[(i)] $C$ is positive semi-definite, i.e.  $ \nu^T C \nu \ge 0 \hspace{5mm} \forall \nu \in V^{(n)} $
\item[(ii)] All diagonal elements of $C$ are one , i.e. $C_{ii} = 1 \hspace{5mm} i=1,\ldots,n$
%\end{enumerate}
\end{itemize}
\end{defin}

\begin{rem}
Since any projection of $C$ onto the subspace of $V^{(n)}$ is positive semi-definite, for any pair $i,j \in\left\{1,\ldots,n\right\}$ the $2\times 2$ matrix $M$ defined by   \newline
\linebreak \vspace{10mm}
$\hspace{10mm} M_{k(\tilde{i}),k(\tilde{j})} = e_{\tilde{i}}^T C e_{\tilde{j}} \hspace{10mm} \tilde{i},\tilde{j} \in \left\{i,j\right\}, k(\tilde{i}) = 1+1_{\tilde{i}=j}$   \newline
is a  correlation matrix as well.
As $det M \geq 0$ it follows that $ -1 \leq C_{\tilde{i} \tilde{j}}  =  M_{k(\tilde{i}) k(\tilde{j})} \leq 1 \hspace{5mm}$.
Hence from Definition \ref{def:corrMatrix} it follows that all off-diagonal elements of $C$ must be between -1 and 1.
\end{rem}

%\begin{figure}
%\label{chol}
%\begin{center}
%\includegraphics[width=0.6\textwidth]{cholesky}
%\caption{: Let x,y,z random variables with corr(x,y) = 1 and corr(x,z) = -1. In this case the corr
%          between y and z cannot be +1. In fact a negative eigenvalue is encountered in
%          this case}
%\end{center}
%\end{figure}
                   
%\begin{rem}  The positive semi-definiteness of C ensures ``causality" between correlated ``movements"  between random variables. To see why consider the movement of three random variables depicted in Fig. 4.  
%\end{rem}

\begin{lem}
\label{lem:lcmcorrel}
Let $V^{(n)}$ be a $n$-dimensional vector space in $R$ and $\rho$ a correlation matrix . Let $u \in V^{(n)}$ be an arbitrary vector and $\kappa \in \{0,1\}$. The matrix defined by
\begin{align}
\label{aaa}
\hat{\rho}_{ij} \equiv
\begin{cases}
\frac{\rho_{ij}+\kappa u_iu_j}{\sqrt{(1+u_i^2)(1+u_j^2)}} \indent{} i \ne j \in \{1,\ldots,n \} \\ \\
\hspace{5mm} 1	 \indent{} \hspace{15mm} i = j
\end{cases}
\end{align}
is a correlation matrix.
\end{lem}

\begin{proof}
In order to prove Lemma \ref{lem:lcmcorrel}, it suffices to construct, for a fixed time $T=1$, $n$ random variables $X_T^i$ $i=1,...,n$ with 
$$0 < E^Q [(X_T^i-E^Q(X_T^i))^2] < \infty \hspace{5mm}and \hspace{5mm}  E^Q [(X_T^i-E^Q(X_T^i))(X_T^j-E^Q(X_T^i))] < \infty $$
such that
\begin{align}
\label{form:cond}
\frac{E^Q[(X_T^i-E^Q(X_T^i))(X_T^j-E^Q(X_T^j))]}{\sqrt{E^Q[(X_T^i-E^Q(X_T^i))^2]E^Q[(X_T^j-E^Q(X_T^j))^2]}}
\end{align}
yields Eq.\ref{aaa}.
Let $(\Omega,F_t,Q)$ be a probability space where the filtration $F_t$ is generated by a $n$-dimensional correlated Brownian motion $\omega_t$ together with two Poisson process $J_1 = N_t^{(1)} \nu_1, J_2 = N_t^{(2)} \nu_2$ of (the same) intensity $\lambda$  where $N_t^{(i)} \hspace{2mm} i=1,2$ is the number of jumps of Poisson process i that occurred between 0 and $t$ and $\nu_i$ is the jump size. $\omega_t$ is given by
\begin{align}
\label{correlated}
d<\omega^i_t,\omega^j_t> = \rho_{ij}(t) \hspace{2mm} dt
\end{align}
If one defines 
$$X_t^i \equiv \omega^i_t-N_t^{(i)} \nu_i$$
It is straightforward to show that
$$E[X_T^i] = -\lambda T \nu_i \hspace{1cm}\text{and}\hspace{1cm} E[(X_T^i)^2] = T + (\nu_i)^2\lambda T (1+\lambda T).$$
In the case where both jumps are triggered simultaneously, e.g. $ N_t^{(1)} =  N_t^{(2)} $ , a direct computation of Eq. \ref{form:cond} yields
\begin{align}
\label{form:corrMatrix1}
\hat{\rho}_{ij} \equiv \frac{\rho_{ij}+\lambda \nu_i \nu_j}{\sqrt{(1+\lambda \nu_i^2)(1+\lambda \nu_j^2)}}
\end{align}
In the case where both jumps are independent one finds  
\begin{align}
\label{form:corrMatrix2}
\hat{\rho}_{ij} \equiv \frac{\rho_{ij}}{\sqrt{(1+\lambda \nu_i^2)(1+\lambda \nu_j^2)}}
\end{align}
Equations \ref{form:corrMatrix1} and \ref{form:corrMatrix2} can be combined with the result
\begin{align}
\label{form:finalCorrMat}
\hat{\rho}_{ij} \equiv \frac{\rho_{ij}+\kappa u_i u_j}{\sqrt{(1+u_i^2)(1+u_j^2)}}
\end{align}
where $\kappa = \left\{ 0,1 \right\}$ and $u_i \equiv \sqrt{\lambda} \nu_i$ $\Box$
\end{proof}
\begin{rem}
Equations \ref{form:corrMatrix1} and \ref{form:corrMatrix2} are consistent with intuition: In cases where $sgn(\nu_i)=sgn(\nu_j) $, correlated jumps increase term correlations between two random variables whereas uncorrelated jumps tend to decrease correlations (in absolute terms). 
\end{rem}
\begin{rem}
One can reduce Equation \ref{form:finalCorrMat} to a 1-parameter family of correlations with parameter  $u \in R$ by introducing a (fixed) ``principal mode" $\xi \in V^{(n)}$ to describe perturbations $u_i = u \hspace{1mm} \xi_i$ of the correlation matrix, e.g.
\begin{align}
\label{form:corrFam}
\hat{\rho}_{ij}(u) \equiv
\begin{cases}
\frac{\rho_{ij}+\kappa u^2\xi_i\xi_j}{\sqrt{(1+\xi^2_i u^2)(1+\xi_j^2 u^2)}} \indent{} i \ne j \in \{1,\ldots,n \} \\ \\
\hspace{5mm} 1	 \indent{} \hspace{22mm} i = j
\end{cases}
\end{align}

%We say that Equation \ref{form:corrFam} defines a 1-parameter family of correlations with center $\rho(u)=(\rho)_{ij} $ $i,j=1,...n$ and mode $\xi$.

This motivates the following

\end{rem}
\begin{defin}
The set
\begin{align}
\label{form:genCorrFam}
 \hspace{0.5cm} 
F_{\kappa}(\rho,\xi) \equiv \left\{ \hat{\rho}_{ij}(u): \hspace{0.2cm} u \in R;\hspace{5mm} \kappa \in \{0,1\};\hspace{0.5cm} \xi \in V^{(n)}; \hspace{0.5cm} \hat{\rho}_{ij}(u) \equiv \frac{\rho_{ij}+\kappa u^2 \xi_i \xi_j}{\sqrt{(1+\xi_i^2 u^2)(1+\xi_j^2 u^2)}} \indent{} i \ne j  \right\}
\end{align}
where $\hat{\rho}_{ii} = 1 $ is called a one-dimensional family of correlation matrices with center $\rho(u)=(\rho)_{ij} $ $i,j=1,...n$ and mode $\xi$.
\end{defin}
\begin{rem}
\label{lcmcorr}
In the special case of a flat mode, i.e. $\xi = (1,1,\ldots,1)^T$ Equation \ref{form:genCorrFam} reduces to
\begin{align}
\label{lcmsimple}
\hat{\rho}_{ij}(u) \equiv 
\begin{cases}
\frac{\rho_{ij} + \kappa u^2}{(1+u^2)} \indent{} i \ne j \in \{1,\ldots,n \} \\ \\
\hspace{5mm} 1	 \indent{} \hspace{5mm} i = j
\end{cases}
\end{align}
\end{rem}

This particular form will prove very useful for the construction of the LCM model because of the analytic tractibility
that goes along with it. This will turn out to be crucial if one wants to construct a model that will be of practical use.

\pagebreak

\section{Local Correlation model (LCM)}
\label{chap:LCM}

The goal of this chapter is to extend the classic multi-asset Dupire local volatility model by introducing the notion of local correlations into the dynamics between the assets. Local correlations imply that, during an infinitesimal time-step, assets evolve according to a correlation matrix that depends on the current state of the system and generally varies from one time-step to another. The main result of this chapter is the derivation of a particular functional form of the correlation matrix that achieves consistency with the index skew by construction. This is crucial if one wants to avoid lengthy fitting procedures of correlation parameters.

\paragraph{}

\begin{assum}
\label{assum:mkt}

Let $(\Omega,F_{t},Q)$ be a probability space where the filtration $F_{t}$ is generated by a
                           n-dimensional correlated Brownian motion $W_t = (\omega^{1}_{t},...,\omega^{n}_{t})$ e.g. $F_{t}= F^{\bold W}$  where

%\begin{align}
%  d< \omega^i_t,\omega^i_t > = \rho_{ij} dt  
%\end{align}

%\begin{align}

%d\< \omega^i_t,\omega^i_t \> = \rho_{ij} dt  

%\end{align}
\begin{align}
d< \omega^i_t,\omega^j_t > = {\rho}_{ij}(t) \;\;\;  dt 
\end{align}

\vspace{1cm}
 $ {\rho}_{ij}(t)$ is a correlation matrix with  $ {\rho}_{ij}(t) \geq 0$. There exist n assets  $S_i(t)$ $i=1,...,n$ that evolve according to the following local volatility
dynamics:

\begin{align}
\label{lognorm}
\frac{dS_t^i}{S_t^i}=\mu_t^i  dt + \sigma_i (t, S_t^i) \;\;  d\omega_t^i \;\; \;\;\;\; i=1,\ldots,n
\end{align}

where $\mu_t^i$ denotes a deterministic drift. The zero index defines a basket

\begin{align}
\label{basketdef}
 S_t^0 \equiv \sum_{i=1}^{n}  \alpha_i S_t^i
\end{align}
together with
\begin{align}
\label{diffbasketdef}
 dS_t^0 = \sum_{i=1}^{n}  \alpha_i dS_t^i
\end{align}

for some given basket weights  $ 0 \leq \alpha^i < \infty $. %where the individual volatilties obey the Novikov condition, i.e.
\vspace{0.5cm}
Locally there exists a Brownian motion $\omega_t^0 $ that matches the dynamics Eq.\ref{diffbasketdef} in a weak sense, e.g.

\begin{align}
\label{lognormbsk}
\frac{dS_t^0}{S_t^0}=\mu_t^0  dt + \sigma_0 (t, S_t^0) d\omega_t^0\;\; 
\end{align}

%$$ E\left[ exp(\frac{1}{2} \int_{0}^{T^*} \sigma_i^2(t,S_t^i)) \right] < \infty $$
\vspace{1cm}

Furthermore for each asset $ i = \{0,...,n\}$ there exists a  market that trades plain vanilla European options at all maturities $T<T^*<\infty$ and strikes $K < K^* \leq \infty $. The set of call options with price $C^i_t(T,K) $ at time t of the i-th asset is denoted by

\begin{align}
\label{optionmkt}
\left\{  C^i_t(T,K) \hspace{5mm} :  \hspace{5mm} T<T^*;\hspace{3mm}K \leq K^*;\hspace{3mm} i=0,...,n \right\}
\end{align}

where 

\begin{align}
\label{continuous}
\frac{\partial C^i_t(T,K)}{\partial T}, \frac{\partial^2 C^i_t(T,K)}{\partial K^2} \hspace{10mm} i=0,...,n 
\end{align}

exist and are continuous. For a single economy $ \large{\varepsilon_i} $ $ (i=0,...,n )$ consisting of a determinstic bond $B(t,T) = exp(- \int_t ^T(r_s \hspace{1mm} ds)) $ together with $ S_t^i  \hspace{3mm} $, i.e. $ \large{\varepsilon_i} =  \{B(t,T),S_t^i\}$, there exists an equivalent martingale measure (EMM) $Q_i$.

\end{assum}

\begin{lem}
\label{thm:dupire}
Under assumption \ref{assum:mkt}, each single option market Eq.\ref{optionmkt} has no arbitrage if the following statements are correct:
\newline

%\begin{align}
(i)\begin{equation}{\label{arb}} 
%$\label{arb}
  \hspace{25mm} \mu_t^i = r_t-q_t^i \hspace{5mm} 
%\end{noalign}
\end{equation}
\newline
\newline
where $r_t$ denotes the (deterministic) instantaneous interest rate and $q_t^i$ the dividend yield. 
\newline
\newline
$ \indent \hspace{0mm}(ii)$
\begin{align}
\label{arb1}
  \sigma_i^2(t,S_t^i) = \frac{\frac{\partial C^i_t(T,K)}{\partial T} + q_t^i C^i_t(T,K)+\mu_t^i K  \frac{\partial{C}^i_t(T,K)}{\partial K}}{\frac{1}{2} K^2 \frac{\partial^2 C^i_t(T,K)}{\partial K^2}} \hspace{10mm}i=0,...,n 
\newline
\newline
\vspace{10mm}
\end{align}
\end{lem}
$$
\linebreak
$$
Lemma \ref{thm:dupire} states the Dupire-equation \cite{Dupire}. As there is some confusion in the literature about the drift terms we repeat the proof in Appendix 1.

\begin{rem}

The lack of continuity condition of type Eq.\ref{continuous} generally leads to serious impairments in the model's ability to  calibrate to the market. Mostly this occurs when market participants define their own internal volatility surface representation that is based on parameterizations that include sharp cut-offs or by kludging together parameterizations of different types at some point of the curve in a non-differentiable way.
\end{rem}

The following lemma deals with cross-arbitrage (also called dispersion-arbitrage ) between the individual option markets and the option market of the basket itself

\begin{lem}
\label{arb2}
Under assumption \ref{assum:mkt} there is no-dispersion arbitrage between individual option markets and the option market of the basket if and only if 
\begin{align}
\label{thm:multidupire}
\newline
\newline
%$ \indent \hspace{0mm}(iii)$
\hspace{25mm} S_t^0 \mu_t^0 = \sum_{i=1}^{n} \alpha_i \mu_t^i S_t^i 
\end{align}
\end{lem}
and for any $t<T^*$ and for any $  (S_t^1,...,S_t^n)\in R^{n,+} $ one has 
\begin{align}
\newline
\label{arb3}
\hspace{20mm} \left(\sum_{i=1}^{n} \alpha_i S_t^i \right)^2 \hspace{2mm} \sigma_0^2(t,S_t^0) = \sum_{i,j=1}^{n} \alpha_i \hspace{1mm} \alpha_j S_t^i \hspace{1mm} S_t^j \sigma_i(t,S_t^i) \hspace{1mm} \sigma_j(t,S_t^j) \hspace{1mm} \rho_{ij}(t)
\end{align}

As the basket weights are assumed greater or equal to zero in Eq.\ref{basketdef}, the maximal covariance is achieved in the case when the correlation matrix is identical one, i.e. $\rho_{ij}=1$ . The lower bound is given in the case when the correlation matrix equals the identity matrix due to the non-negative correlations assumption made in Assumption \ref{assum:mkt}. This gives rise to the following

\begin{cor}
Under assumption \ref{assum:mkt} a necessary condition for no-dispersion arbitrage between individual option markets and the option market of the basket is 
\begin{align}
\label{arb4}
\hspace{20mm} \sum_{i=1}^{n} \left(\alpha_i S_t^i \sigma_i(t,S_t^i) \right)^2 \hspace{2mm} \leq  \left( \sum_{i=1}^{n} \alpha_i S_t^i \right)^2  \sigma_0^2(t,S_t^0) \leq
\sum_{i,j=1}^{n} \alpha_i \hspace{1mm} \alpha_j S_t^i \hspace{1mm} S_t^j \sigma_i(t,S_t^i) \hspace{1mm} \sigma_j(t,S_t^j)
\end{align}
%\end{lem}
\end{cor}

Note that Eq.\ref{arb4} could be violated by the market since all variables are infered from it. If this happens it would indicate arbitrage opportunities provided the form of Eq.\ref{lognorm} is correct.

%\begin{rem}
% Eq.\ref{arb},\ref{arb1} ensure the consistency of the dynamics with the option and forward market of the individual components	while %Eq.\ref{arb2},\ref{arb3},\ref{arb4} ensure consistency with the basket. 
%\end{rem}

%\begin{rem}
%\label{lvlimit}
%Note that for n = 1, Theorem \ref{thm:noarb} reduces to the standard single-asset local volatility model. For zero basket weights $\omega_i = 0 \hspace{3mm} %i=1,...,n$  Equations \ref{arb2}, \ref{arb3}, \ref{arb4} ensuring the consistency of the index with its components are trivially fulfilled and the standard %multi-asset version of the local volatility model is obtained in this case.\footnote{It is worth mentioning that it has become standard for investment banks to have %an implementation of discrete dividends available.}
%\end{rem}

%\begin{rem}
%\label{gatheral}
%In several instances Eq.\ref{arb1} has been published incorrectly. The second term in
%the numerator $ q_t^i \hspace{3mm} C^i_t(T,K) $ is mistakenly replaced by $ \mu_t^i \hspace{3mm} C^i_t(T,K) $.
%\end{rem}
%\ref{thm:multidupire}

\begin{proof}
Note that from Ito's lemma one obtains
\newline
\newline
$ \indent (S_T^0-K)^+  = C_t^0(T,K) + \int_t^T dC_u^0(T,K) $
\newline
\newline
$\indent \hspace{20mm}  = C_t^0(T,K) + \int_t^T {\frac{\partial C_u^0(T,K)}{\partial S_u^0}  (dS_u^0- \mu_u^0 S_u^0 du) } $
\newline
\newline
$ \indent \hspace{24mm} + \int_t^T \frac{1}{2} \frac{\partial^2 C_u^0(T,K)}{\partial^2 S_u^0} d<S_u^0,S_u^0>+\left( \frac{\partial C_u(T,K)}{\partial u}+\frac{\partial C_u^0(T,K)}{\partial S_u^0}  \mu_u^0 S_u^0 \right) du  $
\newline
\newline
$\indent \hspace{20mm}  = C_u^0(T,K) + \int_t^T {\frac{\partial C_u^0(T,K)}{\partial S_u^0}  (dS_u^0- \mu_u^0 S_u^0 du) } $
\newline
\newline
where $ d<S_u^0,S_u^0> =   du \left(\sigma_0(u,S_u^0) S_u ^0 \right)^2 $ which follows from Eq.\ref{lognormbsk}.
\newline
\newline
The cancellation of the last two terms is due to the Black-Scholes partial differential equation but can also be viewed as a direct consequence of the martingale representation theorem. The last line describes the hedge-replication strategy of the option payout in terms of delta-hedging with the basket spot.
In order to prove the absence of dispersion arbitrage one needs to re-write this equation as hedge-integral in terms of its components itself. If one inserts 
Eq.\ref{diffbasketdef} one achieves this goal as long as the drift terms still cancel so that discounted option prices remain martingales. This is indeed the case as long as the quadratic variation of the basket obeys
 
%From Equation \ref{basketdef} and the fact that there is no "spot-arbitrage" between the basket and its constituents (see Appendix 1) it follows that $ dS_u^0 = %\sum_{i=1}^{n}  \alpha_i dS_u^i.$ Note that from assumption (iv) one observes the following identity for the quadratic variation:
%\newline

$$  dt \left(\sigma_0(t,S_t^0) S_t ^0 \right)^2 = d<S_t^0,S_t^0> = dt \sum_{i,j=1}^{n} \alpha_i \hspace{1mm} \alpha_j S_t^i \hspace{1mm} S_t^j \sigma_i(t,S_t^i) \hspace{1mm} \sigma_j(t,S_t^j) \hspace{1mm} \rho_{ij}(t) $$ 
\newline
\newline

Hence the above hedging strategy can be re-written in terms of an arbitrage-free replication strategy that hedges into the individual components instead:
\newline
\newline
$$  (S_T^0-K)^+  = C_t^0(T,K) + \int_t^T {\frac{\partial C_t^0(T,K)}{\partial S_t^0}  (\sum_{i=1}^{n}\omega_i (dS_t^i- \mu_t^i S_t^i dt)) } $$
\newline
%Because of $\omega_i \ge 0 $ and the assumption of $ \rho_{ij} \ge 0 $ , Eq.\ref{arb4} follows immediately from Eq.\ref{arb3}.
$\Box$

\end{proof}

\begin{thm}
\label{thm:lcm}
$\bf{(LCM)}$ 
Let $F_{\kappa}(\rho_0,\xi)$ be a one-dimensional family of correlation matrices with center $\rho_0$ and mode $\xi=(1,...,1)^T $ and $\hat{\rho}_{ij}(u) \in F(\rho_0,\xi) $. Let

\begin{align}
cov_0  \equiv \sum_{i,j=1}^{n} \alpha_i \alpha_j  S_t^i S_t^j   \sigma_i(t,S_t^i) \sigma_j(t,S_t^j)(\rho_0)_{ij}
\end{align}
\begin{align}
cov_1  \equiv \sum_{i,j=1}^{n} \alpha_i \alpha_j  S_t^i S_t^j   \sigma_i(t,S_t^i) \sigma_j(t,S_t^j)
\end{align}
Under assumption \ref{assum:mkt} and the assumption of Eq.\ref{arb4} the option market of  Eq.\ref{optionmkt} has no arbitrage if Eq.\ref{arb},\ref{arb1},\ref{thm:multidupire} hold and the correlation $\rho_{ij}$ in Eq.\ref{arb3} is set to

$$ \rho_{ij}(t)=\hat\rho_{ij}(u^*) $$ where

\begin{align}
\label{ustar}
u ^*  =
\begin{cases}
\sqrt{\left(-\frac{cov_0-\sigma_0^2(t,S_t^0) (S_t^0)^2}{cov_1-\sigma_0^2(t,S_t^0) (S_t^0)^2}\right) } &
\text{if} \hspace{0.5cm} cov_0-\sigma_0^2(t,S_t^0) (S_t^0)^2  < 0 \\  \\
-\sqrt{\left(\frac{cov_0-\sigma_0^2(t,S_t^0)(S_t^0)^2} {\sigma_0^2(t,S_t^0) (S_t^0)^2} \right)}   & \text{else}
\end{cases}
\end{align}

$\kappa$ is set to one when $ cov_0-\sigma_0^2(t,S_t^0) (S_t^0)^2  < 0 $ and zero else.

\end{thm}

\begin{proof}
\label{lcmproof}
The no-arbitrage hypothesis of individual markets $\varepsilon_i$ follows directly from Lemma \ref{thm:dupire}. In order to proof the absence of dispersion-arbitrage, because of Lemma \ref{arb2}, it suffices to construct a correlation $\hat\rho_{ij}$ that satisfies Eq.\ref{arb3} explicitly. If one inserts Eq.\ref{lcmsimple} into Eq.\ref{arb3} one observes the following cases:
If $cov_0 < \sigma_0^2(t,S_t^0) (S_t^0)^2 $ the center correlations are not large enough to satisfy Eq.\ref{arb3} and one needs to set $\kappa = 1 $ in Eq.\ref{lcmsimple} as $\hat\rho_{ij}\rightarrow 1 $ for $ u\rightarrow\infty$. Note that the covariance matrix defined by

\begin{align}
 cov(u^2) \equiv   \sum_{i,j=1}^{n} \omega_i \hspace{1mm} \omega_j S_t^i \hspace{1mm} S_t^j \sigma_i(t,S_t^i) \hspace{1mm} \sigma_j(t,S_t^j) \hspace{1mm} \frac{(\rho_0)_{ij}+\kappa u^2}{1+u^2}
\end{align}

is a strictly increasing function in $u^2$ for $\kappa =1 $ and a solution must exist in this case due to the bound in Eq.\ref{arb4}. The result is given by the first line of Eq.\ref{ustar}.

Similarly if the center correlation is too large implying that $cov_0 > \sigma_0^2(t,S_t^0) (S_t^0)^2 $ one needs to set $\kappa = 0 $ in this case as $\hat\rho_{ij} \rightarrow 0 $ for $ u \rightarrow\infty$ and $i \neq j$. $cov(u^2)$ is a strictly decreasing function and hence invertible in this area with a solution sure to exist due to the lower bound of Eq.\ref{arb4}. The result is given by the second line of Eq.\ref{ustar}. We note here that in the case where non-flat modes are defined (see Eq.\ref{form:corrFam}) similar arguments apply for the existence of solutions to Eq.\ref{arb3} provided $\xi_i > 0$. The explicit solution may require the application of a simple numerical root-finding procedure however $\Box.$ 

\begin{rem}

For simplicity we restricted ourselves to a situation where $\rho_{ij} \ge 0$ since this is generally fulfilled in the equity market. However assume that, instead of the bounds given by Eq.\ref{arb4}, more general bounds are obeyed by the market 

\begin{align}
\label{arb5}
\hspace{10mm} \sum_{i,j=1}^{n} \alpha_i \hspace{1mm} \alpha_j S_t^i \hspace{1mm} S_t^j \sigma_i(t,S_t^i) \hspace{1mm} \sigma_j(t,S_t^j) \rho_{ij}^{down} \hspace{2mm} \leq  \left( \sum_{i=1}^{n} \alpha_i S_t^i \right)^2  \sigma_0^2(t,S_t^0) \leq
\sum_{i,j=1}^{n} \alpha_i \hspace{1mm} \alpha_j S_t^i \hspace{1mm} S_t^j \sigma_i(t,S_t^i) \hspace{1mm} \sigma_j(t,S_t^j)
\end{align}
%\end{lem}

where $\rho_{ij}^{down}$ is a correlation matrix that may contain negative correlations entries. In order to accomodate this case also Eq. \ref{form:corrFam} needs to be generalized according to

\begin{align}
\label{form:corrFam2}
\hat{\rho}_{ij}(u) \equiv
\begin{cases}
\frac{\rho^{down}_{ij}+ u^2\xi_i\xi_j \rho^{up}_{ij}}{\sqrt{(1+\xi^2_i u^2)(1+\xi_j^2 u^2)}} \indent{} i \ne j \in \{1,\ldots,n \} \\ \\
\hspace{5mm} 1	 \indent{} \hspace{22mm} i = j
\end{cases}
\end{align}

This can easily be achieved by correlating not only the Bronian motions by  $\rho^{down}$ but also the changes in the counting processes $ dN_i $ by means of $\rho^{up} $ in the proof of Lemma \ref{lem:lcmcorrel}.

The corresponding equation for $u^*$ are quite similar to before and given by

\begin{align}
\label{ustar3}
u ^*  =
\begin{cases}
\sqrt{\left(-\frac{cov_0-\sigma_0^2(t,S_t^0) (S_t^0)^2}{cov_1-\sigma_0^2(t,S_t^0) (S_t^0)^2}\right) } &
\text{if} \hspace{0.5cm} cov_0-\sigma_0^2(t,S_t^0) (S_t^0)^2  < 0 \\  \\
\sqrt{\left(-\frac{cov_{-1}-\sigma_0^2(t,S_t^0) (S_t^0)^2}{cov_0-\sigma_0^2(t,S_t^0) (S_t^0)^2}\right)}   & \text{else}
\end{cases}
\end{align}

where

\begin{align}
cov_{-1}  \equiv \sum_{i,j=1}^{n} \alpha_i \alpha_j  S_t^i S_t^j   \sigma_i(t,S_t^i) \sigma_j(t,S_t^j)(\rho^{down})_{ij}
\end{align}

\end{rem}

\end{proof}

\pagebreak

\section{Simulation Results of LCM}
\label{chap:simresults}

\begin{figure}
\begin{center}
\includegraphics[width=0.9\textwidth]{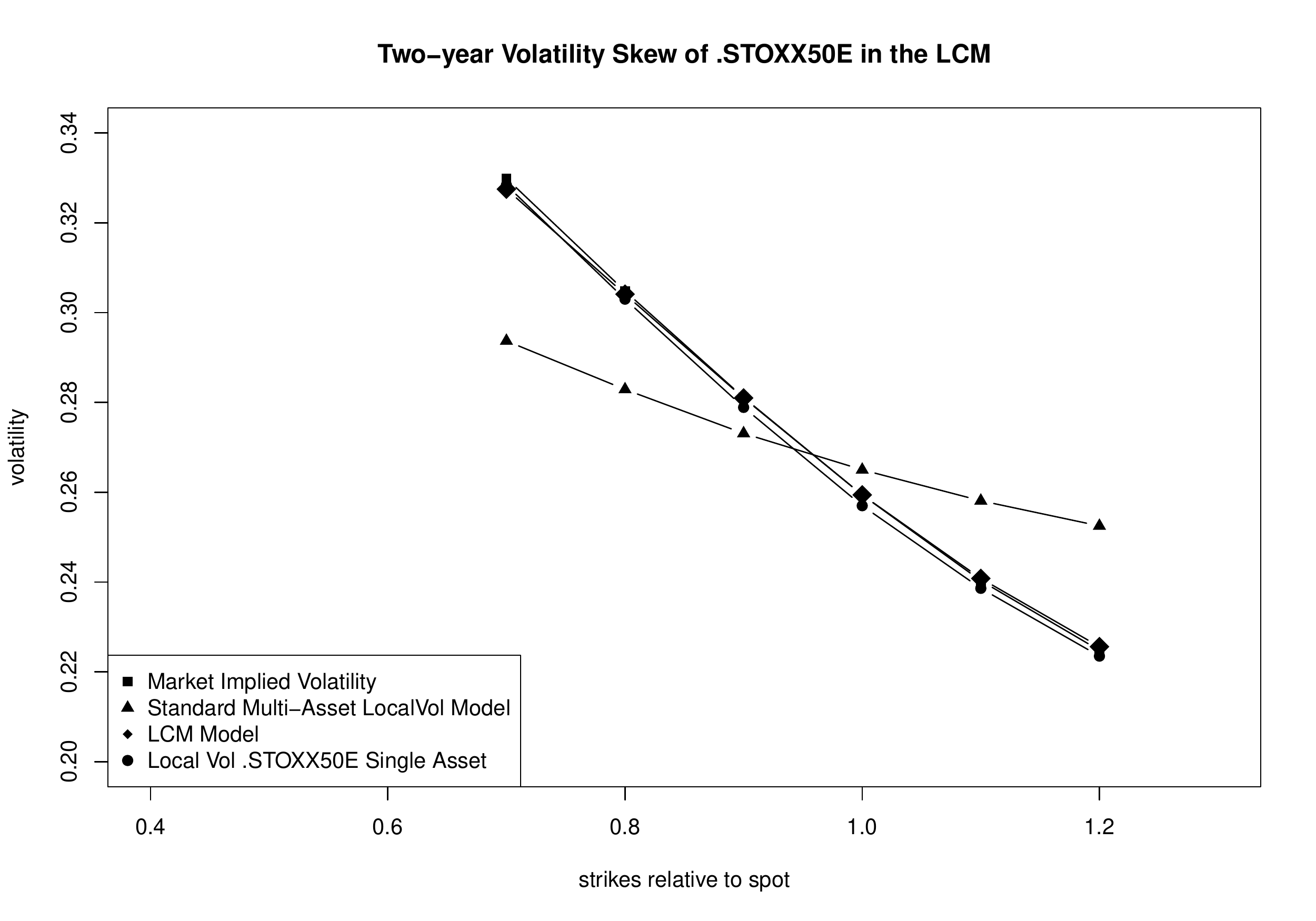}
\caption{The calculation compares the market implied volatility quotes for 2-year options with the (standard) multi-asset local volatility model as well as LCM. Whereas LCM tracks the market distribution quite closely, multi-asset local vol deviates from the market substantially. In order to identify possible sources of numerical errors, we included EuroStoxx 50 in the same simulation as a ``single" asset as well.}
\label{fig:lcmfig}
\end{center}
\end{figure}

In chapter \ref{sect:decoding} we presented the results of the Gaussian copula calculation and demonstrated that the volatility skew of the components of an index do not suffice to explain the traded skew of the index itself. For the results presented here, the choice of correlation matrix was made similar to other market-participants even though the qualitative results are independent of the choice of correlations.

In this chapter we show the results of a similar calculation using the multi-asset local volatility model described in Eq.\ref{lognorm},\label{correlated} when the basket consistency equation \ref{arb3} is not enforced. As one can see in Fig.~\ref{fig:lcmfig} the results are qualitatively similar to the copula calculation of chapter \ref{sect:decoding}: A constant correlation cannot explain the steepness of the volatility skew of the index.

In addition Fig.~\ref{fig:lcmfig} presents results using the LCM model. LCM closely tracks the distribution of the index.

LCM can provide information for the average (off-diagonal) correlations for a given option. The results are presented in Table ~\ref{tab:correlres}. The results show that for a two-year put option struck at $70 \%$ of its spot the correlation is $15 \%$ percentage points higher than for a corresponding call option struck at $120\%$ of spot. To our knowledge this strike-dependence of correlations is substantially higher than what a trading desk typically accounts for in taking reserves for correlation risk (typically $5\%$ points for liquid instruments) . Hence, one expects mispricings in cases where an exotic payout is convex.
The corresponding results for 4-year options are given in table 4.

\begin{table}
\begin{center}
\begin{small} 
\begin{tabular}{|r|r|}
\hline
\hline
Strike in terms of spot &	average correlation in percent\\
\hline
0.7	& 58.3 \\
0.8	& 56.1 \\
0.9	& 53.8  \\
1.0	& 50.6 \\
1.1	& 46.9 \\
1.2	& 43.5 \\	
\hline
\hline
\end{tabular}
\caption{Average instantaneous correlations of two-year options between the constituents of EuroStoxx 50 along the simulated paths in LCM. Correlations are significantly higher on the downside.}
\label{tab:correlres}
\end{small}
\end{center}
\end{table}

\begin{table}
\begin{center}
\begin{small} 
\label{correlres4}
\begin{tabular}{|r|r|}
\hline
\hline
Strike in terms of spot &	average correlation in percent\\
\hline
0.7	& 55.0 \\
0.8	& 51.2  \\
0.9	& 47.7  \\
1	  & 44.5  \\
1.1	& 41.9  \\
1.2	& 39.5  \\
\hline
\hline
\end{tabular}
\caption{Average instantaneous correlations of four-year options between the constituents of DAX along the simulated paths in LCM. Correlations are significantly higher on the downside.}
\end{small}
\end{center}
\end{table}

\pagebreak

\section{Implementation details of LCM}
\label{chap:impldet}

In this chapter we briefly outline how Monte Carlo code needs to be structured to implement LCM.

Consider a simulation of nPaths, nAssets and nDates. Traditionally Monte Carlo code generates one path at a time for each asset in turn.
However this will not work in this case as the choice of correlation matrix depends on the state defined by $all$ the assets. Hence the structure should look as follows:

\vspace{0.3 cm}

\begin{program}

\BEGIN
	\CALL |PrepareCholesky|(numberOfStates, shift, centerCorrelation)
	\FOR ipath:=1 \TO nPaths
		currentPath = initialSpots
		
		\FOR idate:=1 \TO nDates
			randoms = |GenerateRandomNumbers|()
			StateVar = |CalculateCorrelationState|(currentPath)
			CorrelatedRandoms = |CorrelateRandoms|(randoms, CholeskyDecomp[StateVar])
			
			\FOR iasset:=1 \TO nAssets
				PathArray[ipath][iasset][idate] = |GeneratePathArray|(CorrelatedRandoms)
			\END
				currentPath = PathArray[ipath]
		\END
	\END
	
	\CALL |CalcPayout|(PathArray)
\END

\end{program}

In order to correlate random numbers one needs to perform a Cholesky decomposition. In the case where correlations are constant the latter needs to be computed just once, prior to the simulation of the paths. This is not true for LCM as every set of paths will generally require a different matrix to be decomposed. In order not to adversely impact performance it is useful to prepare a table of correlation matrices beforehand together with their Cholesky decompositions. The table implements a discrete version of $F(\rho_{base}, \xi =(1,...,1)^T) $ for $\hat{\rho}_{ij}(u^*)$ from Eq. \ref{lcmsimple} for about 50-100 values for $u^* $ , along the lines of:

\begin{program}

\BEGIN{PrepareCholesky(numberOfStates, shift, centerCorrelation):}
	\COMMENT{This function prepares a list of an odd number of correlation}
	\COMMENT{matrices together with its Cholesky decomposition and stores the}
	\COMMENT{results together with the state variable. The index k = k* = (numberOfStates - 1) / 2}
	\COMMENT{corresponds to centerCorrelation. For k $>$ k* one sets $\kappa=1$ in Eq.\ref{ustar}}
	\COMMENT{whereas for  k $<$ k*  it is set to zero in this case.}

	\IF numberOfStates = even
		numberOfStates := numberOfStates + 1
	\END
	
	m := (numberOfStates-1) / 2
	correl[m] = centerCorrelation
	
	\FOR k:=m+1 \TO numberStates
		l := k - m
		|Sum Over i,j|
			\IF i = j
				correl[i][j] = 1
			\ELSE
				u = shift + l * shift
				correl[k][i][j] = correl[k][j][i] = (centerCorrelation[i][j] + u^2) / (1 + u^2)
				StateVariable[k] = u
				cholesky[k] = CholeskyDecompose(correl[k])
				correl[m-l][i][j] = correl[m-l][j][i] = centerCorrelation[i][j] / (1 + u^2)
				StateVariable[m-l] = -u
				cholesky[m-l] = CholeskyDecompose(correl[m-l])
			\END
	\END
\END
\end{program}

Once the simulation has started the function $CalculateCorrelationState(currentPath)$	simply calculates the state variable according to Eq.\ref{ustar} and looks up the closest position in the pre-calculated table from $PrepareCholesky$. Note that all current path values together with each constituent's local volatility and the local volatility of the index need to be available to compute Eq.\ref{ustar}.

\pagebreak

\section{Residual (exotic) Correlation Risks in the Pricing of Multi-asset Derivatives: The ``Chewing-Gum" Effect}
\label{chap:chewinggum}

LCM allows one to compute single stock delta and vega exposures for multi-asset derivatives. In addition it produces sensitivities when the basket volatility skew and thus the distribution of the basket is changing. As the basket volatility depends on the correlation, vega hedging of the index component of an exotic, to some extent, hedges the correlation exposure as well.
\newline
\newline
In a sense the idea is to ``lock in" the projection of the exotic instrument onto the basket ``subspace" and therefore to hedge some of the correlation exposure along with it.

However a hedge into the index only hedges overall (correlation) movements of the market. Mathematically, the specification of the distributions of the sum together with its components does not suffice to fully describe the joint distribution.
We give a simple example in footnote \cite{example}.

In order to demonstrate this point, consider a basket $B_t$ of two equally weighted stocks $S_1(t),S_2(t)$  with $S_1(0)=100,S_2(0)=100$  , e.g.

$$ B_t = \frac{1}{2} (S_1(t)+S_2(t)) $$

and consider a ``chewing-gum" move where  $S_1(t)$ goes to 80 and 
 $S_2(t)$  to 120. As the basket level stays the same in such moves, neither the individual distributions nor the basket distribution change in this case. However an option that pays the worst of the two assets is affected in this case.
\newline

In this chapter we try to get an estimate of this residual risk by finding alternative solutions that provide consistency with the basket skew. Any price sensitivity of worst-of options will be attributed to exotic correlation risk in this case.

The LCM model was introduced for a family of correlations $\hat{\rho}_{ij} \in F(\rho,\xi)$ centered around $\rho$  with a flat mode $\xi=(1,...,1)^T$. The merits of this choice lie in the fact that $\hat{\rho}(u)$ can easily be inverted analytically in this case with the result given by Eq.\ref{ustar}. However the model can easily be extended to non-flat perturbation provided $\xi_i > 0 \hspace{1mm}\forall \hspace{1mm}{i=1,...,n}$ . This is feasible as long as one is willing to deal with a slightly more involved condition for the inversion of $\hat{\rho}(u)$ (see remark at the end of the proof of theorem \ref{thm:lcm}).
\newline
In this paper we pursue a simpler approach by simply varying the center-correlation instead. Recall that the center-correlation is merely a starting point around which a table of correlation matrices is constructed together with their Cholesky decompositions. The individual selection of correlation matrices is done locally according to Eq. \ref{ustar}.
\newline
\newline
In the following we present the results for two-year put options on the worst-of all 50 assets of EuroStoxx 50, which pays at maturity 
\newline
$$ (K-Min(\frac{\small{S_1(T)}}{\small{S_1(0)}},...,\frac{S_{50}(T)}{S_{50}(0)}))^+ $$

for two different choices for the center-correlation:

Set 1 presents results based on $F(\rho,(1,...,1)^T)$ where $\rho$  was chosen close to where we believe the ``market" was trading whereas set 2 is based on $F(diag(1,...,1),(1,...,1)^T)$ with the identity matrix chosen as center.

\begin{table}\begin{small} 
\begin{tabular}{|r|r|r|}
\hline
\hline
Strike in terms of spot &	worst-of put price: set1 &	worst-of put price: set2\\
\hline
0.6	& 0.306 & 0.463\\
0.7	& 0.391 &  0.560\\
0.8	& 0.482 & 0.657\\
0.9	& 0.575 & 0.754\\
\hline
\hline
\end{tabular}
\caption{Price of worst-of put}
\label{tab:worstoff}
\end{small}
\end{table}

We have also computed the average correlation which comes out roughly the same in both cases. Henceforth despite the fact that all individual as well as basket distributions are the same, the value of worst-of puts substantially differ between the two sets. In a sense basket options are mainly sensitive to overall shifts of the correlation matrix only and hence to the first principal component of the correlation matrix itself. This is different for worst(best)- of call and put options which show significant sensitivity to ``higher" vibrations such as chewing-gum moves representing higher principal components of the correlation matrix. This is why liquid prices for worst(best)-of options would be very useful in providing further important clues about the joint dynamics of the system. In other words, the volatility skew for the basket can provide only little insight into the pricing of the worst-of (best-of) options, particularly when the number of basket members is large. This makes them ideal candidates to augment the correlation information provided by the basket.

\pagebreak

\section{Conclusion}

In this paper we show that the index option's market is inconsistent with the option market of the individual constituents if
the dependency between the assets was to be described by a deterministic correlation matrix only or by a Gaussian copula.
The data strongly suggests higher correlations on the downside and gives rise to a correlation skew which is computed explicitly
in this paper. Roughly only half of the skew of the index can be attributed to the skew of the individual components, indicating that state-dependent correlation dynamics could play a crucial role in explaining the distribution of the index.

This paper generalizes Dupire's local volatility model to the multi-asset case in a numerically efficient way and provides consistency
not only with the individual option's markets but also with the volatility market of the index itself by making correlations a dynamic variable.

We also discuss the non-uniqueness of the solution and introduce the "`chewing-gum"' effect in this context where single stocks move against each other while leaving the distribution of the basket as well as all individual distributions unchanged. However worst-of options show price sensitivity in such moves which makes them ideal candidates to further ``complete" the market and augment our knowledge of the joint dynamics of the system.

\pagebreak

\section{Appendix 1}

In the first step consider $n+1$ separate economies $ \large{\varepsilon_i} $ for $ i=0,...,n $ each one consisting of a deterministic bond $B(t,T) = exp(- \int_t ^T(r_s \hspace{1mm} ds)) $ together with $ S_t^i  \hspace{3mm} $, e.g. $ \large{\varepsilon_i} =  \{B(t,T),S_t^i\}$.
Because of the specific form of Eq.\ref{lognorm}, that is the fact that volatility depends on the spot only, each $ \large{\varepsilon_i} $  is complete. The existence of an equivalent martingale measure states that the dividend-adjusted processes $\tilde{S}_t^i = S_t^i/e^{-q t}$ are martingales under the cash numeraire $B(t,T)$  yielding Eq.\ref{arb} \cite{harrison}. The specific form of Eq.\ref{arb1} has been derived in the literature already (see for example \cite{Dupire}). However this paper ignores drift terms. In several instances Eq.\ref{arb1} has been published incorrectly. The second term in
the numerator $ q_t^i \hspace{3mm} C^i_t(T,K) $ is mistakenly replaced by $ \mu_t^i \hspace{3mm} C^i_t(T,K) $. We quickly re-derive this equation at the end of this Appendix. 
In the second step one considers the economy $ \large{\varepsilon}=\{ B(t,T),S_t^0,...,,S_t^n \} $ that trades the option
market specified in Eq.\ref{optionmkt}. In order to avoid ``spot" arbitrage between the index and its components taking appropriate $(Q_i$ )� expectation values on both sides of Eq.\ref{basketdef} immediately yields Eq.\ref{thm:multidupire}. One is left to show that Eq.\ref{arb3} avoids dispersion arbitrage between the basket option market and the individual option markets.
This is done in Chapter \ref{chap:LCM} .
\newline
\newline
Re-derivation of Dupire's formula:

If one defines

$$ \xi_t \equiv (S_t-K)^+ \beta_t $$ where $ \beta_t \equiv exp(-\int_0^t {r_s ds}) $. The application of Tanaka's formula yields
$$ d\xi_t = \beta_t \left( 1_{S_t>K} dS_t + \frac{1}{2}\delta(S_t-K)^+ S_t^2 \sigma_t^2 dt - r_t (S_t-K)^+ \right) $$

Stochastic integration on both sides with subsequent expectation taking yields:

$$ C_0(T,K) - (S_t-K)^+ = \int_0^T {dt \hspace{1mm} \beta_t \left( \mu_t E[S_t 1_{S_t>K}]+\frac{1}{2} K^2 E[\delta(S_t-K) \sigma_t^2]-r_t E[S_t-K)^+]\right)}$$
Differentiation on both sides with respect to T gives:
$$ \frac{\partial C_0(T,K)}{\partial T}  = \beta_T \left( \mu_T \hspace{1mm} E[S_T 1_{S_T>K}]+\frac{1}{2} K^2 E[\delta(S_T-K) \sigma_T^2]-r_T E[S_T-K)^+]\right)$$

Note that plain vanilla options obey

$$ C_0(T,K) = \beta_T \left( E[S_T 1_{S_T>K}]-K E[1_{S_T>K}] \right) $$
The second expectation is just a call spread, e.g. $ E[1_{S_T>K}] = -\frac{\partial C_0(T,K)}{\partial K} $.
Similarly the expectation of the Dirac delta function can be inferred from option prices according to $ E[\delta[S_T-K)^+ ] = \frac{\partial^2 C_0(T,K)}{\partial^2 K} $.
If one inserts this back into the above equation and uses the fact that
$$ E[\delta(S_T-K) \sigma_T^2] =  E[\delta(S_T-K)] E[\sigma_T^2 \|S_T=K] $$
this yields Eq.\ref{arb1}, as for a state-dependent local vol the expectation drops, e.g. $E[\sigma_T^2 \|S_T=K]=\sigma_T^2(S_T=K) $

\pagebreak

\section{Acknowledgement}

I would like to thank Romain Barc for many inspiring discussions on the subject. I am greatful to Alexander Kabanov, Daniel Cangemi and Thilo
Meyer-Brandis for useful comments. I also would like to thank Michael Demuth, Thomas Quillet and Syed Uzair Aqeel for valuable suggestions on the paper.

%\cite{Langnau2007}

\bibliographystyle{plain}

\begin{thebibliography}{}

\end{thebibliography}


\begin{thebibliography}{100}
\bibitem[1]{Langnau2006} \emph{Option pricing when correlations are stochastic: an analytical framework}, Jos�e Da Fonseca, ESILV �and Zeliade Systems, September 2006.
\bibitem[2]{cit1} \emph{Explaining the index skew via stochastic correlation models}, Matthias R. Fengler, Helmut Herwartz , Christian Menn , Christian Werner.
\bibitem[3]{cit2} \emph{Option-Implied Correlations and the Price of Correlation Risk}, Joost Driessen, Pascal Maenhout, Grigory Vilkov, March 2005.
\bibitem[4]{cit3} \emph{The Shape and Term Structure of the Index Option Smirk:Why Multifactor Stochastic Volatility Models Work so Well}, Peter Christo�ersen Steven Heston Kris Jacobs.
\bibitem[5]{cit4} \emph{Towards a Generalization of Dupire's Equation for Several Assets}, P. Amster , P. de Napoli  and J. P. Zubelli, April 2006.

\bibitem[6]{harrison}\emph{"Martingales and Stochastic integrals in the theory of continuous trading". Stochastic Processes and their Applications}, Harrison, J. Michael; Pliska, Stanley R. (1981). 

\bibitem[7]{Dupire}\emph{Pricing and hedging with smiles}, Bruno Dupire,April 1993.

\bibitem[8]{example}{Let $\omega_1,\omega_2,\omega_3$ be three independent Brownian Motions. Consider the set 1 of two variables $ X_1 = \omega_1+\omega_2+\omega_3$ and $ X_2 = -\omega_1 $ as well as set 2 containing  $ X_3 = \omega_1+\omega_2+\omega_3$ and $ X_4 = -\frac{1}{\sqrt{2}}\omega_1-\frac{1}{\sqrt{2}}\omega_2$. Even though individually the variables share the same individual distributions between the two sets together with the distribution of the sum, the joint distributions differ between the two sets.}

\end{thebibliography}

\end{document}